\documentclass[apjl,twocolumn]{emulateapj_mod}
\usepackage{epsfig,apjfonts,mathptmx}

\def\gtsima{$\; \buildrel > \over \sim \;$}
\def\ltsima{$\; \buildrel < \over \sim \;$}
\def\prosima{$\; \buildrel \propto \over \sim \;$}
\def\gsim{\lower.5ex\hbox{\gtsima}}
\def\lsim{\lower.5ex\hbox{\ltsima}}
\def\simgt{\lower.5ex\hbox{\gtsima}}
\def\simlt{\lower.5ex\hbox{\ltsima}}
\def\simpr{\lower.5ex\hbox{\prosima}}

\def\h1{$h^{-1}$}
\def\eeq{\end{equation}}
\def\beq{\begin{equation}}
\def\24mu{24\,$\mu{\rm m}$}
\def\70mu{70\,$\mu{\rm m}$}
\def\8mu{8\,$\mu{\rm m}$}

\shorttitle{The mass-SFR correlation in $z=1$ galaxies}
\shortauthors{Salmi et al.}
\begin{document}

\title{Dissecting the stellar mass-SFR correlation in \lowercase{$z=1$} star-forming disk galaxies}

\author{
F. Salmi\altaffilmark{1},
E. Daddi\altaffilmark{1},
D. Elbaz\altaffilmark{1},
M. T. Sargent\altaffilmark{1},
M. Dickinson\altaffilmark{2},
A. Renzini\altaffilmark{3},
M. Bethermin\altaffilmark{1},
D. Le Borgne\altaffilmark{4}
}

\altaffiltext{1}{Laboratoire AIM, CEA/DSM - CNRS - Universit\'e Paris Diderot,
       Irfu/Service d'Astrophysique, CEA Saclay, Orme des Merisiers,  91191 Gif-sur-Yvette Cedex, France}
\altaffiltext{2}{National Optical Astronomy Observatory,  950 N. Cherry Ave., Tucson, AZ 85719}
\altaffiltext{3}{INAF-Osservatorio Astronomico di Padova, Vicolo dell'Osservatorio 2, I-35122 Padova, Italy}
\altaffiltext{4}{Institut d'Astrophysique de Paris, UMR 7095 CNRS, UPMC, 98 bis boulevard Arago, 75014, Paris, France}

\email{fadia.salmi@cea.fr, edaddi@cea.fr}
 
\begin{abstract}
Using a mass-limited sample of 24$\mu$m-detected, star-forming galaxies at $0.5<z<1.3$, we study the mass-star formation rate (SFR)
correlation and its tightness. The correlation is well defined
($\sigma=0.28$~dex)
for disk galaxies ($n_{\rm sersic}<1.5$), while more bulge-dominated objects often have lower specific SFRs.
For disk galaxies,
a much tighter correlation ($\sigma=0.19$~dex) 
is obtained if the rest-frame H-band luminosity is used instead of stellar mass derived from multicolor photometry. The specific SFR (sSFR) correlates strongly 
with rest-frame optical colors (hence luminosity-weighted stellar age) and also with clumpiness 
(which likely reflects the molecular gas fraction). 
This implies that most of the observed scatter is real, despite its
low level, and not dominated by random measurement errors. 
After correcting for these differential effects
a remarkably small dispersion remains ($\sigma=0.14$~dex), suggesting that measurement errors
in mass or SFR are $\simlt0.10$~dex, excluding systematic uncertainties. 
Measurement errors in stellar masses, the thickening of the correlation due to real sSFR variations, and varying completeness with stellar mass,
can spuriously bias
the derived slope to lower values due to the finite range over which observables (mass and SFR) are available. When accounting for these effects,
the intrinsic slope for the main sequence for disk galaxies gets closer to unity.
\end{abstract}


\section{Introduction}

Star-forming galaxies obey a tight correlation between stellar
mass and star formation rate (SFR),
 from the local Universe (e.g., Elbaz~et~al.\ 2007; based on
 Brinchmann~et~al.\ 2004; Peng~et~al.\ 2010), all the way  to
intermediate and at high redshifts  (e.g., Noeske~et~al.\ 2007; Elbaz~et~al.\ 2007; Daddi~et~al.\ 2007; 
Pannella~et~al.\ 2009a; Magdis~et~al.\ 2010; Karim~et~al.\ 2011; Daddi~et~al.\ 2009; Stark~et~al.\ 2009; 
Lee~et~al.\ 2011; Gonzalez~et~al.\ 2010; Bouwens~et~al.\ 2011). 
With a scatter of $\sim 0.3$ dex at all redshifts where it has been measured, 
this relation is now known as the {\it Main Sequence} (MS) of star-forming galaxies.
This finding has several interesting implications: (1) there is a high degree
of uniformity among star-forming galaxies, and stellar mass is a crucial parameter regulating the SFR; 
(2) fluctuations in specific SFRs throughout the star formation histories of actively star-forming
galaxies are minor in all but
a small number of 
outliers (e.g., Rodighiero~et~al.\ 2011); (3) at high redshifts, the SFR of individual galaxies must increase rapidly with
time (Daddi~et~al.\ 2007; Renzini 2009; Peng~et~al.\ 2010; Papovich~et~al.\ 2011). 

The existence of such a tight relation  raises several questions.
First, it is often believed that stellar masses and SFRs for individual galaxies cannot be
measured to much better than a factor of two precision, even in relative
terms. 
This raises the question whether the small spread in the mass-SFR correlation is actually dominated
by measurement
errors, implying a potentially smaller intrinsic scatter. The question  applies to other tight
relations recently discovered for normal galaxies, like that between
mid-IR and total IR luminosity (Elbaz~et~al.\ 2011, scatter $\sim 0.3$ dex),
and between SFR and CO luminosity (scatter  $\sim 0.2$ dex; Daddi~et~al.\ 2010ab; Genzel~et~al.\ 2010; Sargent~et~al.\ 2012, in preparation). 
Intimately related to this is a second issue which deals with the slope of 
the correlation. Writing SFR$\propto M_*^{\alpha}$, it is generally
found that $\alpha\simlt1$, but results vary from $\sim 0.6$ to $\sim1$, 
depending on sample definition, the adopted SFR indicator, and (perhaps) redshift
(e.g.,  Pannella~et~al.\ 2009a; Karim~et~al.\ 2011). 
Finally, it is generally found that the mass-SFR correlation holds 
for star-forming galaxies only, but the term ``star-forming galaxy'' is somewhat ambiguous.
For example, one can refer to star-forming BzK samples at $z\sim2$ 
(e.g., Daddi~et~al.\ 2007; Pannella~et~al.\ 2009a), or to blue-cloud galaxies at $z=0$ to 1 (e.g., Elbaz~et~al.\ 2007; Peng~et~al.\ 2010), or to Lyman 
break galaxies at $z\geq3$.  It would be desirable to explicitly study what happens for objectively selected
and complete samples of all star-forming galaxies down to low levels of specific SFR (henceforth, sSFR), 
well below the MS.  To start addressing these questions, we will explore in this letter
the role of physical parameters, including morphology, in driving the mass-SFR correlation.
We assume a WMAP concordance cosmology and a Chabrier IMF.

\begin{figure}[ht]
\centering
\includegraphics[width=8.8cm]{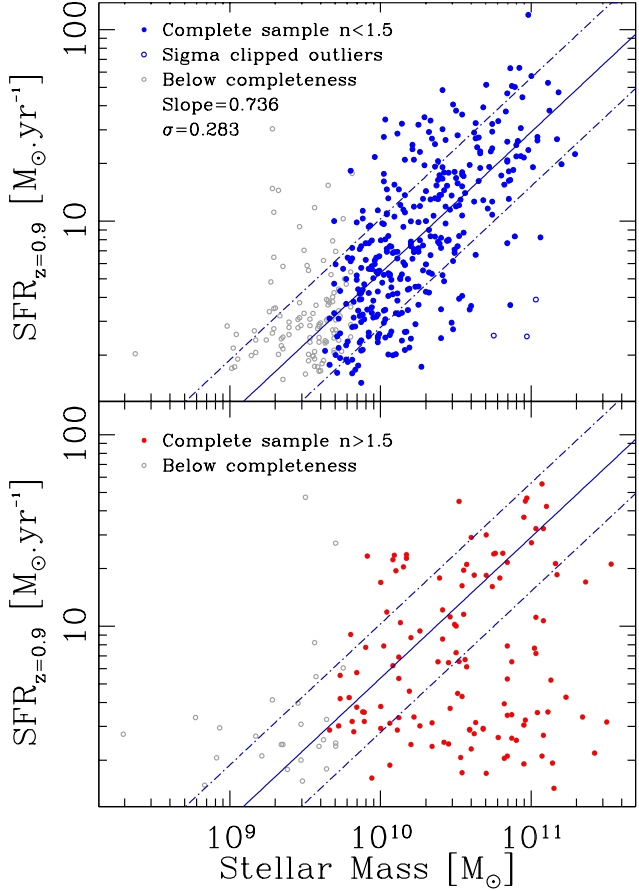}
\caption{The stellar mass-SFR correlation for disk galaxies ($n_{\rm sersic}<1.5$, top) and bulge dominated galaxies ($n_{\rm
sersic}>1.5$, bottom). 
Lines in the top panels show
the best fit relation and the $1\sigma$ scatter after 3-sigma clipping of outliers, and are repeated in the bottom panel for comparison.
Points below completeness are not used in the fit.
}
\label{fig:1}
\end{figure}

\section{Sample selection and measurements}
\label{sample}

We study a sample of galaxies at $0.5<z<1.3$ extracted from the
$K$-band selected catalog of Daddi~et~al.\ (2007) in GOODS-S. 
We consider all star-forming galaxies with a S/N$>3$ detection and
flux $>12\mu$Jy at $24\mu$m from Spitzer+MIPS (e.g., Magnelli~et~al.\ 2011). 
We remove 56 AGNs significantly detected in the 2~Ms Chandra X-ray data (Alexander~et~al.\ 2003).
There are 599 galaxies in the sample, of which 70\% have spectroscopic
redshifts and the remainder have accurate photometric redshifts 
from Grazian~et~al.\ (2006). Detailed spectral energy distributions (SEDs)
are available based on multi-color imaging from the $U$-band through the mid-infrared, which we use to derive rest frame magnitudes and colors by 
spline-interpolation between adjacent photometry. Stellar
masses are derived from SED fitting ($M_*^{\rm SED}$ hereafter), using a wide range of star formation histories, metallicity, and
allowing for dust reddening, using the method described in Le~Borgne \& Rocca-Volmerange (2002) and as used in Elbaz et al.\ (2007; 2011).
The Spitzer 24$\mu$m fluxes are converted into SFRs using SED templates from Chary \& Elbaz (2001).  Analyses of Spitzer and Herschel far-infrared data have shown this approach to be
reliable for galaxies at $z < 1.3$ (Magnelli et al.\ 2009;  Elbaz et al.\ 2010, 2011).  The infrared-based SFR is added to the unobscured component computed from the UV 1500\AA\
luminosity (extrapolated from the observed photometry using the best-fitting SED model), without correction for dust extinction.
By comparing the latter to the total
SFR, we obtain a direct estimate of the dust attenuation at 1500\AA\ (A$_{1500}$) that we use to correct absolute magnitudes 
and rest frame colors for reddening,  based on the Calzetti~et~al.\ (2000) extinction law. 
These extinction-corrected magnitudes and colors are used throughout
the following analysis.

\begin{figure}[ht]
\centering
\includegraphics[width=8.8cm]{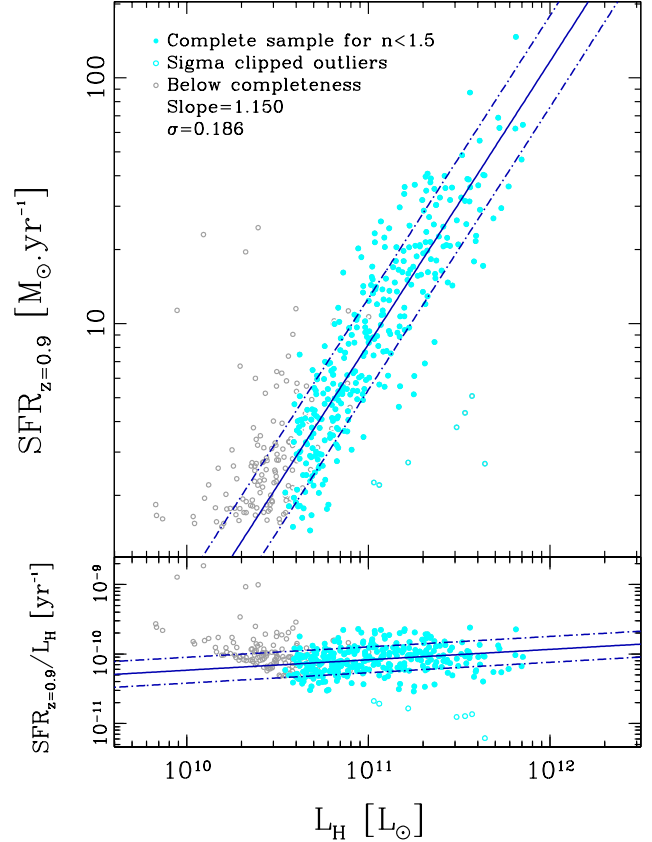}
\caption{
Correlation between SFR and H-band luminosity. 
} 
\label{fig:2}
\end{figure}

One aspect that we want to explore is the effect of galaxy morphology on the mass-SFR
correlation of star-forming galaxies.
Using {\em GALFIT} (Peng~et~al.\ 2002) we model the HST+ACS z-band
images of each object (Giavalisco~et~al.\ 2004,  release 2.0),
deriving the S\'ersic index ($n_{\rm Sersic}$) and half-light radius. 
It is well known that star-forming galaxies become clumpier at higher redshifts (e.g., Elmegreen~et~al.\ 2004;
Forster-Schreiber~et~al.\ 2009), hence we have also investigated the
role of clumpiness. Galaxy clumpiness is generally defined as the fraction of 
light  in high spatial frequency structures  (Conselice 2003; Lotz~et~al.\ 2004). 
Here we measure
clumpiness ($S$) after subtracting the {\em GALFIT} model ($G$) from the image ($I$) of each galaxy:

\beq
S = \, \langle \frac{|I-G|}{G} \rangle.
\end{equation}
 
\noindent
The images in both the numerator and denominator are convolved by the PSF, and
we exclude pixels within
$\pm1.5\sigma$ to reduce the noise, thus  measuring residual structures only above this intensity threshold. 
The average is performed over the segmentation map of the galaxy, as defined by SExtractor, which is meant 
to produce an {\em intensive} measurement of clumpiness, normalized by the
spatial extent of the galaxy (A$_{\rm S}$). We would expect that this measurement should correlate better with SFR per unit
surface area rather than with SFR itself, and we thus also use an {\em extensive} measurement of clumpiness defined as $S^{\rm ext}=S\times
{\rm A}_{\rm S}$ that should be more closely related to the SFR. We find that $S^{\rm ext}$ is better correlated with visual clumpiness classification 
performed for our sample.
Clumpiness is sensitive to all sorts
of structures, including clumps but also to the strengths of spiral
arms and anything deviating
from a single S\'ersic fit.

\begin{figure}[ht]
\centering
\includegraphics[width=8.8cm]{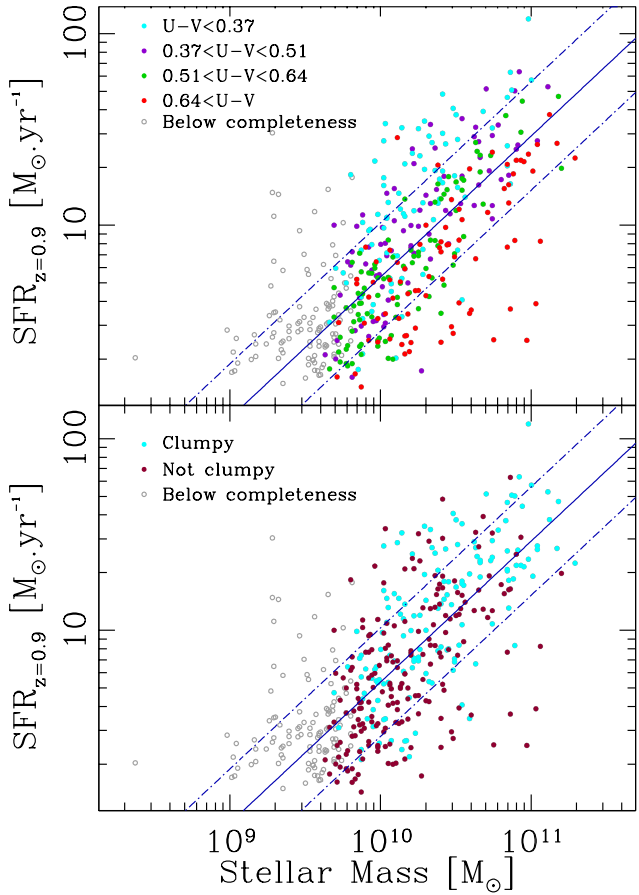}
\caption{
The mass-SFR correlation with galaxies coded according to their optical rest-frame reddening-corrected colors (top) and clumpiness (bottom). Lines 
from Fig.~1-top.
}
\label{fig:3}
\end{figure}

\begin{figure}[ht]
\centering
\includegraphics[width=8.8cm]{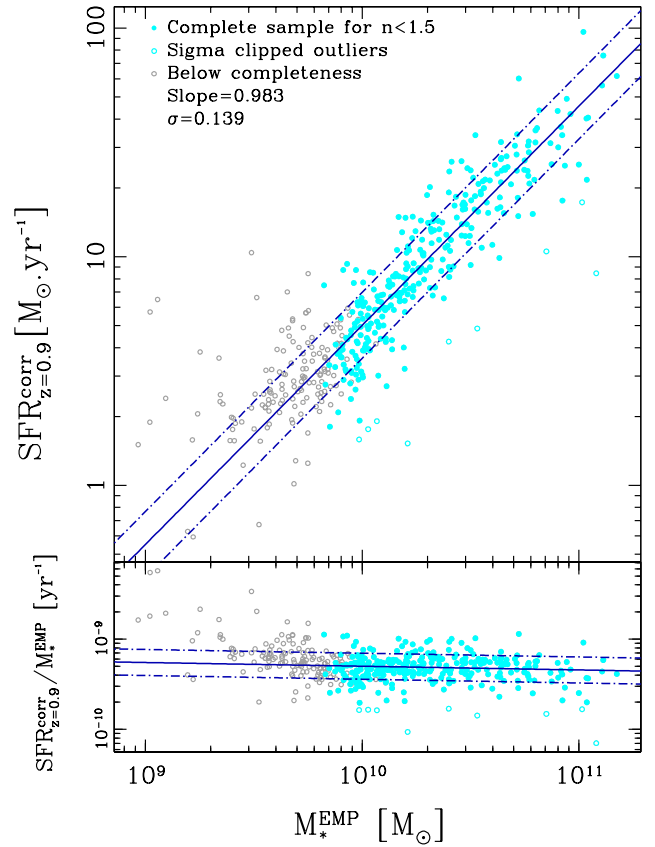}
\caption{
The mass-SFR correlation when SFR is formally corrected for trends in redshift, color and clumpiness (see Table~1). Lines show
the best fit relation and the $1\sigma$ scatter after 3-sigma clipping of outliers.
}
\label{fig:4}
\end{figure}

\newpage

\section{Results} 

When plotting galaxies in the mass-SFR plane (Figure~\ref{fig:1}), 
we find that bona-fide disk galaxies (447 objects with $n_{\rm Sersic}<1.5$, or 72\% of the parent 24$\mu$m-selected sample)
display a clear, well-defined correlation.  
We adopt here a lower threshold than the canonical $n_{\rm Sersic}=2$ separation of disks/ellipticals in order to have a cleaner sample of disks,
at the expense of completeness.
Instead, some more bulge-dominated  ($n_{\rm Sersic}>1.5$) 
galaxies fall near the correlation for disk galaxies, while others
have lower sSFRs (Figure~\ref{fig:1}), 
consistent with the results of Wuyts~et~al.\ (2011). The latter galaxies are often
visually classified as ellipticals and have red optical colors. 
We conclude that the mass-SFR correlation is primarily a sequence of star-forming  disk galaxies.
Comparing to color selections, 96\% of $n_{\rm Sersic}<1.5$ galaxies fall in the blue-cloud,
while conversely 73\% of blue-cloud galaxies have $n_{\rm
  Sersic}<1.5$, implying that a blue-clould selection of star-forming galaxies results
in a higher dispersion.
Including $n_{\rm Sersic}>1.5$ galaxies  would also bias the slope, as completeness as a function of sSFR
varies strongly with mass in a SFR-limited sample (we are considering only 24$\mu$m-detected galaxies), i.e., a Malmquist bias. 
Red, bulge-dominated objects  with sSFR more than 0.6 dex below the MS  contribute $\simlt 6\%$ 
of the total SFR density from $M_*^{\rm SED}>5\times 10^{10}M\odot$ galaxies.
This is likely to be an overestimate because the 24$\mu$m emission in
some of these red spheroids, rather than from star formation,  
could come from winds in evolved stars (e.g., Vega~et~al.\ 2010) or to
an AGN torus, whereas the rejection of X-ray-detected AGNs is 
strongly reducing the number of main sequence galaxies at the highest stellar masses (Mullaney~et~al.\ 2012).
In the following analysis, we consider only the sample of 338 $n_{\rm
  Sersic}<1.5$  disk galaxies satisfying redshift dependent mass
limits which guarantee completeness down to the lower edge of the MS.
At stellar masses higher than such mass limits, galaxies can be detected to even lower sSFRs. 
This can lead to a spurious flattening of the measured slope.

\begin{table*}
{\tiny
\caption{Summary of multilinear fit results for log(SFR) based on Eq.~2. Each row shows a particular fit with the physical parameters used for that fit in column~1,
the relative coefficients determined in the fit (with their S/N between parenthesis) in column~2, the rms between individual SFR measurements and those predicted by the multilinear fit
in column~3, and the ope of the logarithmic SFR-$M_*$ (or SFR-$L_{\rm  H}$) relation in column~4.}
\label{tab:1}
\centering
\begin{tabular}{l|l|c|c}
\hline\hline
Parameters~($P_i$) & Coefficients~($c_i$) & RMS & Outliers~rej. \\
           & Values (S/N)  & dex & Fraction (\%) \\
	   (1) & (2) & (3) & (4)\\

   \hline
$c_\circ$, log M$_*^{SED}$, log ((1+z)/1.9)                                       & -6.631(15.3), 0.736(11.4), 2.791(9.1) & 0.283 & 0.89\\
$c_\circ$, log $L_H$, log ((1+z)/1.9)                                             & -11.733(29.0), 1.150(31.4), 0.689(3.0) & 0.186 & 2.20 \\
$c_\circ$, log M$_*^{EMP}$, log ((1+z)/1.9)                                       & -9.110(18.4), 0.979(20.4), 1.416(4.8) & 0.246 & 1.68 \\
$c_\circ$, log M$_*^{EMP}$, log ((1+z)/1.9), $(U-V)_{\rm rest}$                   & -10.147(32.4), 1.080(35.6), 0.093(0.48), -0.923(21.7)& 0.154 & 2.02 \\
$c_\circ$, log M$_*^{SED}$, log ((1+z)/1.9), $(U-V)_{\rm rest}$                   & -7.329(21.9), 0.804(24.7), 1.632(6.64), -0.969(16.2)& 0.215 & 2.36 \\
$c_\circ$, log M$_*^{EMP}$, log ((1+z)/1.9), $(U-V)_{\rm rest}$, log S$^{\rm ext}$ & -9.145(28.8), 0.983(32.0), 0.406(2.2), -0.956(24.4), 0.149(6.4) & 0.139 & 3.03 \\
Const, log M$_*^{SED}$, log ((1+z)/1.9), $(U-V)_{\rm rest}$, log S$^{\rm ext}$ & -6.345(18.7), 0.709(21.4), 1.917(8.0), -0.944(16.6), 0.224(7.2) & 0.208 & 1.47 \\
\hline
\hline
\end{tabular}\\
}
\centerline{Notes: M$_*^{SED}$, $L_H$ and M$_*^{EMP}$ are in solar Units; the absolute scale of $S^{\rm ext}$ is arbitrary}
\end{table*}

We now explore to which extent real variations of galaxy properties contribute to the thickness of the mass-SFR relation.
We treat SFR as the dependent variable and present multiple linear
fits of log(SFR) as a function of log~mass and other measured galaxy properties ($P_i$'s):

\beq
{\rm log(SFR)} = c_0 +  c_1\times P_1 + ... c_n\times P_n
\eeq

\noindent where $c_i$'s are coefficient to be determined by the fit (see Table~1 for the different fits performed).
Given the fairly wide
redshift range $0.5<z<1.3$ explored, over which the normalization of
the mass-SFR relation changes appreciably, we scale
all SFR measurements to a common $\langle z \rangle = 0.9$ using the relation:\\

${\rm log(SFR_{z=0.9}) = log(SFR) - c_z\times log \frac{(1+z)}{(1+0.9)}}$\\

\noindent
where the coefficient $c_z$ is also derived from the multilinear
regression (Table~1).
By main sequence `slope'  we mean the coefficient of log($M_*)$ (or equivalent) in the
fit, whereas the scatter refers to the rms of the difference between
the individual measures of log(SFR) and that predicted by the
multilinear fit of Eq.~2.
We apply an iterative 3$\sigma$ clipping to remove strong outliers in
the fit, which results in rejecting from $\sim 1$ to at most $\sim 3\%$ of the
galaxies.

The mass-SFR correlation for our disk sample is found to have a slope of 0.74
and a scatter of 0.283~dex (Figure~1). This is just slightly shallower than reported by Elbaz~et~al.\ (2007) but steeper than that of Noeske~et~al.\ (2007)
at similar redshifts, and comparable to the slope and scatter found by Rodighiero~et~al.\ (2011) for $z\sim2$ galaxies.

The rest-frame 1.6$\mu$m luminosity ($L_{\rm H}$) is sometimes used as a proxy for the
stellar mass (e.g., Gavazzi, Pierini \& Boselli 1996; Cowie \& Barger 2008), so we also
tried using this in the fitting.
Remarkably, the SFR-$L_{\rm H}$ relation is even tighter,
with a scatter of only 0.186~dex (Table 1; Figure~\ref{fig:2}).
The $M_*/L_{\rm H}$ ratio is affected by the light from young stars, 
and hence it is somewhat affected by the SFR. However,
$L_{\rm U}$ is even more directly affected by star formation, and yet we find
that the SFR-$L_{\rm U}$  relation has a scatter of 0.40~dex.
When using galaxy colors $(U-V)_{\rm rest}$ together with $L_{\rm H}$
in the fit one obtains a relation with even smaller scatter
(0.154~dex), hence 
colors seem to be the strongest factor affecting the mass-SFR sequence. Binning in colors produces roughly parallel sequences
in the mass-SFR plane (Figure~\ref{fig:3}; top). 
The $M_*/L_{\rm H}$ ratio is also strongly affected by color (e.g., Pannella et
al.\ 2009b):
redder galaxies have lower sSFR and higher $M_*/L_{\rm H}$, which helps reduce the scatter in the $M_*-L_{\rm H}$ correlation 
compared to the SFR-$M_*$ relation. 
We thus introduce
an `empirical mass' $M_*^{\rm EMP}$ by fitting the $M_*^{\rm SED}$ as a function of $L_{\rm H}$ and $(U-V)_{\rm rest}$
color:

\beq
{\rm log}\ M_*^{\rm EMP} = -1.399+1.042\times {\rm log}(L_{\rm H})+0.339\times (U-V)_{\rm rest},
\eeq

\noindent
similar in concept to Eq.~6--7 of Daddi~et~al.\ (2004) for BzK galaxies and to Eq.~1 in Bell (2008) for local galaxies  
(but recall that here quantities are reddening corrected). 
The dispersion between $M_*^{\rm EMP}$ and $M_*^{\rm SED}$ is only 0.20~dex.
We find that the $M_*^{\rm EMP}$-SFR correlation
has a scatter of 0.246~dex, slightly smaller than the scatter in $M_*^{\rm SED}$-SFR  (0.283~dex, Figure 1).
A simple interpretation of this finding would be that, somewhat surprisingly, the `empirical mass' is a more accurate stellar mass estimator by about 0.14~dex in relative
terms (subtracting in quadrature), compared to  using full SED fitting with a large number of galaxy population synthesis models.
This would be possible if the large number of degrees of freedom (on star formation histories,
etc) in the libraries of fitted templates acts to increase the uncertainties through the various degeneracies of actual galaxies, 
perhaps because actual galaxies have more homogeneous star formation
histories than are permitted in the SED-fitting models, leading to a
smaller range of $M_*/L$ ratios (at a given color)
in the real world compared to the models, while the simple fit in Eq.~2 could be effectively equivalent to introducing priors in the SED-selection.
Demonstrating this hypothesis would require extensive simulations that are beyond the scope of this work.
We emphasize that Equation 2 has been empirically calibrated for our  galaxy sample with an effective $z = 0.9$.  It is not immediately obvious if the same relation could be applied at
much lower or higher redshifts.

We also find that clumpiness correlates with the SFR residuals from the average relation:
clumpy galaxies tend to have higher sSFRs (Figure~\ref{fig:3}-bottom; Table~1),
an effect which is significant at $>6\sigma$ (Table~1) when the regression includes
$S^{\rm ext}$ together with reddening-corrected color. The latter fit has a scatter of
0.139~dex (Table~1; Figure~\ref{fig:4}) and a slope of 0.98, which is substantially steeper than the original fit. 
We also tried using $n_{\rm Sersic}$ and  size but these parameters do
not appear to have a measurable correlation with sSFR in our sample of
disk galaxies.

\section{Discussion} 

When color and clumpiness are included as fitting parameters together with stellar mass (Figure~4; Table~1), 
the resulting correlation ultimately reaches a scatter of about 0.14~dex, 
compared to 0.28 for the mass-SFR correlation (Figure~1).
This implies that most  ($\simgt0.24$~dex) of the scatter in the original correlation is real, i.e., due to real variations of sSFR that can be traced to galaxy observables (namely, color
and clumpiness). The combined effect of measurement errors on $M_*$ and SFR is thus at the level of $\simlt0.14$~dex, and likely smaller, as
our two parameters $(U-V)_{\rm rest}$ and $S^{\rm ext}$ may not account for all of the physical variance. If the measurement errors in $M_*$ and SFR are similar, 
then each of these quantities is
precise to better than 0.10~dex in relative terms, or about 25\% in linear scale, which is quite amazing.
It could be argued, however,  that  what we are comparing here is not 
a direct measurement of these quantities. For example, our SFR is
derived from the total infrared luminosity $L_{\rm bol}$, which in
turn is inferred from 12$\mu$m rest-frame luminosity
$L_{12}$ (the UV contribution is generally negligible), hence our conclusions strictly apply to such an
observable, and similar considerations could held for $M_*$. 
In principle, the exact conversion between $L_{12}$ and the bolometric luminosity ($L_{\rm bol}$) could affect the
slope that we infer for the mass-SFR relation, although our adopted value is perfectly consistent with all observables at $z\sim1$ (Elbaz~et~al.\ 2011).
It could also be possible that this $L_{12}$ to $L_{\rm bol}$ conversion actually reduces the observed scatter, so that a smaller rms is found for the tracer (e.g. $L_{12}$)
with respect to what it is designed to trace (e.g., SFR or $L_{\rm
  bol}$). This might be the case if the correlation between IR8
(=$L_{8\mu{\rm m}}/L_{\rm bol}$) and sSFR 
(Elbaz~et~al.\ 2011) were to hold also at 12$\mu$m rest frame and inside the main sequence. On the other hand, for moderately
star forming galaxies mid-IR might be better correlated to SFR than $L_{\rm bol}$ (see, e.g., Calzetti et al.\ 2007)

It is worth commenting on the reason for the dependence of sSFR on color and clumpiness. For the latter the interpretation appears to be 
quite straightforward given that a higher gas fraction is expected to be the reason for higher clumpiness (Bournaud, Elmegreen \& Elmegreen 2007;  Ceverino, Dekel \& Bournaud 2010),
and the higher gas fraction should directly imply higher SFRs (all
other things being equal, hence higher sSFR). Higher sSFR obviously results in
bluer colors, but age, metallicity, or extinction effects may also be
at play. We explored if color differences could be connected to a metallicity 
dependence of the mass-SFR relation (Mannucci~et~al.\ 2010;
Lara-Lopez~et~al.\ 2010). However, the effect of a metallicity-SFR
relation should depend strongly on stellar mass, so it seems unlikely that it could produce the series 
of almost parallel sequences
as a function of color seen in Figure~\ref{fig:3}. In addition, metallicity has a weak impact on optical 
colors of actively star-forming galaxies. Using synthetic galaxy
spectra we find that higher metallicity actually can imply slightly
bluer colors for young galaxies. 
We exclude the possibility that colors reflect, even in part, dust reddening, 
because the luminosities and colors used in this analysis have all been corrected for dust extinction (see Section~2).
We also find that using A$_{1500}$ as an extra parameter would further reduce the scatter to 0.125~dex. 
However, this may be due to the fact that A$_{1500}$ is derived in part from the same infrared data 
used to determine the SFR.

We have demonstrated that a sizable fraction of the dispersion of the
SFR-$M_*$ relation for main sequence galaxies in our sample can be traced to
intrinsic differences in color and clumpiness at fixed mass, as SFR
correlates with these quantities.  While the existence of a main
sequence of star-forming galaxies indicates that star formation
proceeds in a quasi-steady fashion in most galaxies, it is quite
natural to expect that the sSFR is also subject to up and
down fluctuations, possibly responding to fluctuations in the gas
accretion rate. On the other hand, the star formation process is
intrinsically stocastic, as is the formation of clumps with enhanced
SFR.  In addition, some galaxies may sustain systematically lower or higher SFRs, over times
comparable to the Hubble time, which would produce quite diverging 
mass growth histories over cosmological timescales (Renzini~2009).
Addressing this question in detail is beyond the scope of this letter.

Finally, this work provides hints about the possible intrinsic slope of the mass-SFR correlation.
Indeed, the slope on the stellar mass term rises from 0.74 to $\sim1$
when including color and clumpiness or using the empirical mass in the
fit
(see Table~1). 
When the analysis is limited to sSFR ranges with high completeness, the average color of galaxies in our sample does not change with stellar mass, which disfavors the slope steepening 
being just the effect 
of spurious removal of mass-SFR trends. Instead, using color at least partially removes the bias due to the higher dynamic range in sSFR at higher masses, allowing to see galaxies with
lower sSFR and redder colors. 
Also, fitting for color and clumpiness removes most of the (real) thickness of the mass-SFR relation, which can also artificially reduce the slope, as we are spanning only a
relatively small range in both stellar mass and SFR, i.e., about 1.5~dex in this and in similar studies. The slope will also be biased by additional errors
in the stellar mass, which is used here as the independent variable but is also affected by measurement errors. 
Therefore, our work suggests that the intrinsic slope of the mass-SFR
correlation is probably closer to unity than is suggested by simple
fits neglecting these biases.


We conclude by emphasizing that the H-band luminosity and the (reddening corrected) 
rest-frame $(U-V)_{\rm rest}$ color can be used to
predict the 24$\mu$m emission (a measure of dust and PAH luminosity)
with better than a 40\% accuracy,
which is an impressive demonstration of 
the deep connection between the amount of stars already present (the stellar mass) and its time derivative (the star formation rate).

\begin{acknowledgements}
We thank Maurilio Pannella, Samir Salim and the anonymous referee for interesting discussions and suggestions.
FS, ED, MTS, MB  acknowledge funding support from ERC-StG grant UPGAL
240039 and ANR-08-JCJC-0008.
AR acknowledges support from grants INAF-PRIN/2008 and ASI I/009/1-/0.
Based on observations with
the Spitzer Space Telescope, which is operated by the JPL,
CalTech under a contract with NASA. 
\end{acknowledgements}

\end{document}